\documentclass[prd,a4paper,showpacs, nofootinbib, preprintnumbers,amsmath, multicol]{revtex4}

\usepackage{epsfig}
\usepackage{amssymb}
\usepackage{graphicx}

\usepackage{color}
\definecolor{rosso}{cmyk}{0,1,1,0.4}
\definecolor{rossos}{cmyk}{0,1,1,0.55}
\definecolor{rossoc}{cmyk}{0,1,1,0.2}
\definecolor{blu}{cmyk}{1,1,0,0.3}
\definecolor{blus}{cmyk}{1,1,0,0.6}
\definecolor{bluc}{cmyk}{1,1,0,0.1}
\definecolor{verde}{cmyk}{0.92,0,0.59,0.25}
\definecolor{verdec}{cmyk}{0.92,0,0.59,0.15}
\definecolor{verdes}{cmyk}{0.92,0,0.59,0.4}

\def\circa#1{\,\raise.3ex\hbox{$#1$\kern-.75em\lower1ex\hbox{$\sim$}}\,}

\makeatletter
\def\art{\@ifnextchar[{\eart}{\oart}}
\def\eart[#1]#2#3#4#5#6{{\rm #2}, {\em #3 \rm #4} {\rm (#6) #5 ({\em #1})}}
\def\hepart[#1]#2{{\rm #2, \em#1}}
\newcommand{\oart}[5]{{\rm #1}, {\em #2 \rm #3} {\rm (#5) #4}}

\newcommand{\beq}{\begin{equation}}
\newcommand{\eeq}{\end{equation}}
\newcommand{\bea}{\begin{eqnarray}}
\newcommand{\eea}{\end{eqnarray}}
\newcommand{\ba}{\begin{array}}
\newcommand{\ea}{\end{array}}
\newcommand{\bi}{\begin{itemize}}
\newcommand{\ei}{\end{itemize}}
\newcommand{\bn}{\begin{enumerate}}
\newcommand{\en}{\end{enumerate}}
\newcommand{\bc}{\begin{center}}
\newcommand{\ec}{\end{center}}

\newcommand{\gsim}{\lower.7ex\hbox{$\;\stackrel{\textstyle>}{\sim}\;$}}
\newcommand{\lsim}{\lower.7ex\hbox{$\;\stackrel{\textstyle<}{\sim}\;$}}

\begin{document}
\setcounter{page}{0}


\title{Detecting the Cold Spot as a Void with the Non-Diagonal Two-Point Function}

\author{ ~\\   Isabella Masina}  
\email{masina@fe.infn.it}
\affiliation{  {\it Dip.~di Fisica dell'Universit\`a degli Studi di Ferrara and INFN Sez.~di Ferrara, \\
Via Saragat 1, I-44100 Ferrara, Italy  }}
\affiliation{ {\it CP3-Origins, IFK and IMADA, University of Southern Denmark,\\
Campusvej 55, DK-5230, Odense M, Denmark} }
\author{ \vskip -.3 cm and \\~   Alessio Notari}   
\email{alessio.notari@cern.ch}
\affiliation
{\it
Institut f\"ur Theoretische Physik\\
Universit\"at Heidelberg \\
Philosophenweg 16, D-69120 Heidelberg, Germany.
}


\begin{abstract}
~~~\\

{\bf Abstract:}  The anomaly in the Cosmic Microwave Background known as the ``Cold Spot" could be due
to the existence of an anomalously large spherical (few hundreds ${\rm Mpc}/h$ radius) underdense region, 
called a ``Void'' for short. 
Such a structure would have an impact on the CMB also at high multipoles $\ell$ through Lensing. 
This would then represent a unique signature of a Void. 
Modeling such an underdensity with an LTB metric,
we show that the Lensing effect leads to a large signal 
in the non-diagonal two-point function, centered in the direction of the Cold Spot, such that
the Planck satellite will be able to confirm or rule out the Void explanation for the Cold Spot, for {\it any} Void radius 
with a Signal-to-Noise ratio of at least ${\cal O}(10)$.

\end{abstract}


\baselineskip=13pt
\setcounter{page}{1}


\maketitle

\vskip 0.5cm
\section{Introduction}

One of the Cosmic Microwave Background (CMB) anomalies identified in the WMAP~\cite{WMAP} data
is the so-called Cold Spot~\cite{ColdSpot1,ColdSpot2}: 
a spherical region on an angular scale of about $10^\circ$ that appears to be anomalously cold
and whose probability to come from a flat spectrum of Gaussian primordial fluctuations is 
estimated to be about $1\%-2\%$. 
While this could still be due to a statistical fluke (or a fortuitous choice of using a particular 
basis of weight functions\cite{Zhang:2009qg}), 
some authors \cite{Tomita, InoueSilk} have put forward 
the idea that it could be due to an anomalously large underdense region of some unknown origin - 
called a ``Void" for short - located on the line-of-sight between us and the Last Scattering Surface (LSS).
Subsequent papers~\cite{Granett, HuntSarkar, InoueTomita} have shown that other Voids of ${\cal O}(100)$ Mpc/$h$ 
radius seem to be detected, via the correlation between CMB and galaxy surveys, through the Integrated Sachs-Wolfe 
(ISW) effect, and that this would be at odds with the concordance $\Lambda$CDM model. 
Possible inflationary mechanisms which produce these objects involve nucleation of spherical bubbles~\cite{Afshordi}.
As for the direction of the Cold Spot, a claim for a Void at $z \le 1$ based on NVSS radio source data~\cite{rudnick} 
has however been subsequently challenged \cite{huterer, Granett:2009aw, Bremer:2010jn}.

Modeling such an underdensity with a Lema\^itre-Tolman-Bondi (LTB) metric, in two previous papers~\cite{MN, MN2} 
we explored some observational consequences of the hypothesis that a Void is responsible for the Cold Spot 
- see also Ref.~\cite{spergel} for 
a quite similar analysis. 
Traveling through a Void, photons are redshifted due to the fact that the gravitational potential is not 
exactly constant in time, the so-called Rees-Sciama (RS) effect~\cite{ReesSciama}. 
In \cite{MN} we computed the RS effect on the CMB two-point (power spectrum) and three-point (bispectrum) 
correlation functions, which would be affected at low $\ell$.
In \cite{MN2} we showed that, through Lensing (namely the deflection that occurs to a CMB photon 
traveling through a Void), the Void would affect the CMB power spectrum and bispectrum 
also at high $\ell$. We emphasized that this would constitute a unique signature of a Void.
In particular, we found that: for the power spectrum the effect 
will be visible by the Planck satellite for Void radii $L\gtrsim 500~{\rm Mpc}/h$;
for the bispectrum, a signal should be detected by Planck if $L \gtrsim 300~{\rm Mpc}/h$.

In the present paper we extend the analysis of~\cite{MN2} by considering the non-zero correlations in 
the non-diagonal two-point function due to the Lensing effect. Note that the kind of observable 
we consider here is not invariant under rotations but depends on a preferred axis, chosen as the $\hat{z}$ axis 
in the decomposition in spherical harmonics. In this case, however, we know what this preferred axis is, 
since it is exactly the one directed towards the centre of the Cold Spot.
We stress again that such an effect is only present if there actually is a Void on the line of sight, 
while it would be absent if the Cold Spot were just a statistical fluke of the primordial large-scale fluctuations. 
A signal in the CMB non-diagonal two-point function would thus represent a unique signature of a Void. 
As we are going to show, the study of such observable will allow the Planck satellite 
to rule out or confirm the Void explanation of the Cold Spot. 
Moreover the Lensing effect is correlated with the RS effect, which can be seen in some observables, 
such as the three-point correlation function \cite{MN}.

The paper is organized as follows. 
In section II we briefly review the physical effects of an underdense region on the CMB. 
In section III we define a non-diagonal two-point function and compute its Signal-to-Noise ratio.
Finally, we draw our conclusions in section IV.



\section{A Void in the line of sight: Rees-Sciama and Lensing effects}
\label{profiles}

Consider an observer looking at the CMB through a spherical Void with comoving radius $L$ 
and negative density contrast, parameterized by its value at the centre $\delta_0$.
The Void is located at comoving distance $D$ from us, in the direction of the $\hat z$ axis. 
We assume that it does not intersect the LSS and that we are not inside it.
The angle subtended by the Void is $2 \theta_L$, with $\tan \theta_L = L/D$.

The observer receives from the LSS the primordial CMB photons, whose fluctuations we assume to be adiabatic, 
nearly-scale invariant and Gaussian. We also assume that the location of the Void in the sky is not correlated 
with the primordial temperature fluctuations, which is true, for example, if such a structure comes from a 
different process, such as nucleation of bubbles.  
For simplicity we disregard here the effect of a cosmological constant, already considered in \cite{MN2}.

As in~\cite{MN, MN2}, we model the Void's inhomogeneous region via a spherically symmetric LTB metric, 
matched to a Friedmann-Lema\^itre-Robertson-Walker (FLRW) flat model.
From the matching conditions, it follows that our density profile is "compensated", 
{\it i.e.} the underdense central region is surrounded by a thinner overdense shell.
Photons traveling outside the LTB region will not be lensed.

The observer detects one particular realization of the primordial Gaussian perturbations on the LSS {\it plus} 
the secondary effects due to this anomalous structure: the RS redshift effect\footnote{As far as we know, 
the name "Rees-Sciama effect" is generically used when the redshift 
of a photon is due to the non-linear evolution of the gravitational potential, 
as opposed to the name "Integrated Sachs-Wolfe effect" which is usually employed to refer to the evolution of the
potentials already at the linear level, as happens in the presence of a cosmological constant.} ~\cite{MN} 
and the Lensing effect on the photon direction~\cite{MN2}. 
The observed temperature fluctuation is then a sum of three components:
\beq
\frac{\Delta T({\bf \hat n})}{T}~=~
\frac{\Delta T({\bf \hat n})}{T}^{(P)} +~
\frac{\Delta T({\bf \hat n})}{T}^{(RS)} +~
\frac{\Delta T({\bf \hat n})}{T}^{(L)} \, ,
\label{3temp}
\eeq
where (P) stands for primordial, (RS) for Rees-Sciama and (L) for Lensing. 
Each fluctuation is defined as 
$\frac{\Delta T({\bf \hat n})}{T}^{(i)}\equiv ~\frac{T^{(i)}({\bf \hat n})-\bar{T}^{(i)}}{T}$ where $i=P,RS,L$ and 
the bar represents the angular average over the sky and $T=\sum_i \bar{T}^{(i)}=2.73 K$.

A detailed explanation of how to compute the shape for $\Delta T^{(RS)}/T$ can be found in \cite{MN}. 
The RS temperature fluctuation is effectively described by two parameters: 
its amplitude at the centre of the Void, $A=\Delta T({\bf \hat z})^{(RS)}/T$, and its angular extension, 
{\it i.e.} the diameter of the cold region, $\sigma$. Clearly, $\sigma$ is slightly smaller than $2 \theta_L$, 
the angle subtended by the full LTB region. We fix the numerical values of $A$ and $\sigma$ phenomenologically, 
relying on the values given by~\cite{texture}: 
for the temperature at the centre we use the range $T=-(190\pm 80) \mu K$, 
which means $A= (7 \pm 3) \times 10^{-5}$;
for the angular size $\sigma$ of the cold region, we choose the particular but representative values $6^\circ$,
$10^\circ$ and $18^\circ$, which correspond respectively to $\theta_L=7^\circ,11^\circ,20.5^\circ$.

\begin{figure}[h!]\begin{center}
\includegraphics[width=6.7cm]{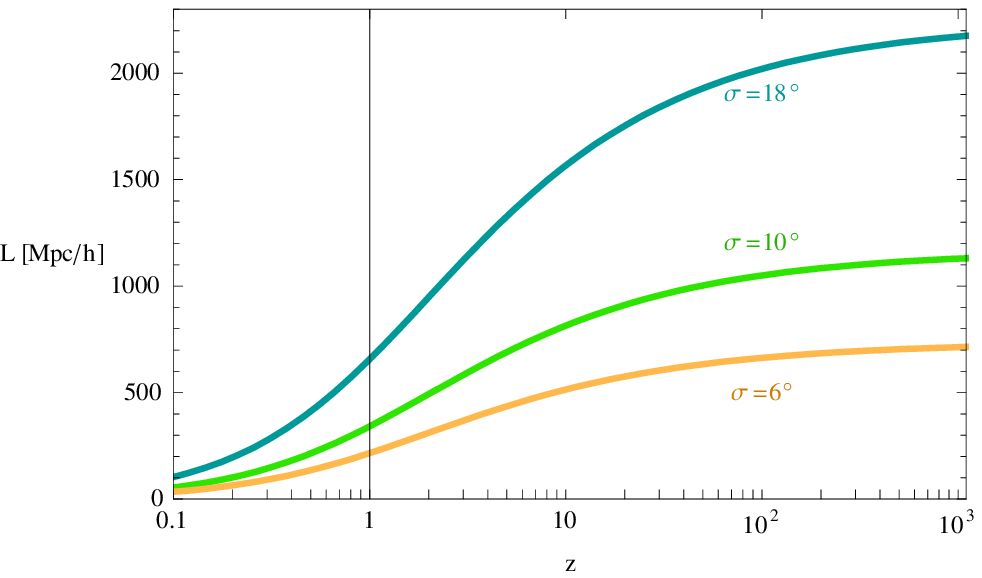} ~~~~~~~~~~~ 
\includegraphics[width=6cm]{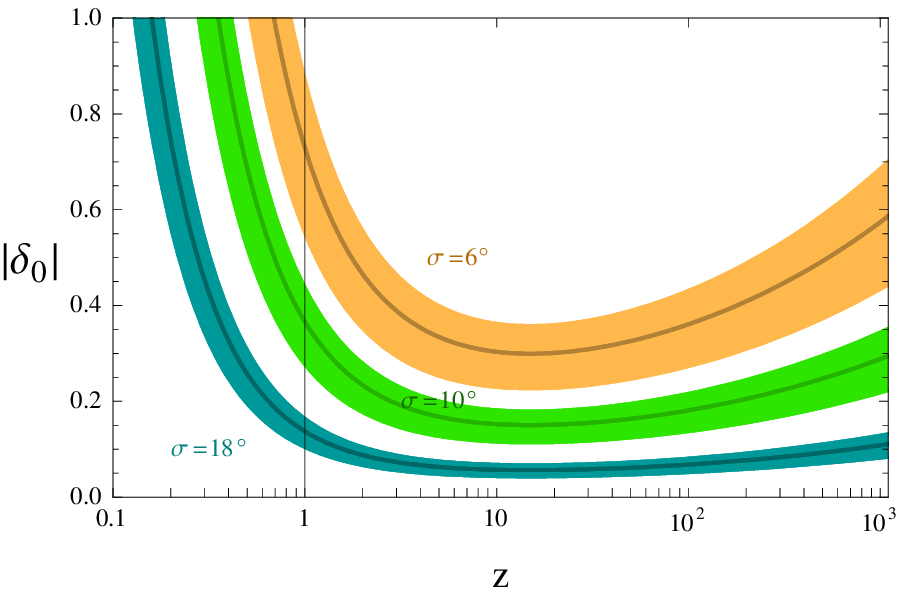}
\end{center}\vspace*{-0.5cm} 
\caption{Plots of $L$ and $\delta_0$ as a function of the redshift $z$, for $\sigma=6^\circ,10^\circ,18^\circ$. 
The shaded regions are obtained by varying $A$ in the range $(7\pm 3) \times 10^{-5}$.}
\vskip .5cm
\label{figLz} 
\end{figure}

We recall from~\cite{MN} that 
we can express $\delta_0$ 
as: 
\beq
|\delta_0| \approx \sqrt{  \frac{2 A}{ 1-\frac{L H_0}{2 \tan \theta_L}} }  ~(L H_0)^{-3/2}~~~.
\label{eqdelta0}
\eeq
It is easy to switch from the dependence on $L$ to the one on 
the redshift at the centre of the Void, $z$, because of the following relation (obtained assuming 
approximately straight lines for photon trajectories): 
\beq
1- \frac{L H_0}{2 \tan \theta_L} = \frac{1}{\sqrt{1+z}}~~.
\eeq 
In fig. \ref{figLz} we show the dependence of $L$ and $\delta_0$ on the redshift $z$,
for the values of $\sigma$ relevant for the Cold Spot.

The temperature fluctuation due to Lensing, $\Delta T^{(L)}/T$, is usually computed in a gradient 
expansion~\cite{Bernardeau, Lewis} and it is given as in~\cite{MN2} once the so-called Lensing potential is known.
This is an integral along the line-of-sight
related to the gravitational potential $\Phi$ as follows:
\beq
\nabla_{\perp}\Theta= -2 \int_{\tau_{LSS}}^{\tau_O} d\tau \frac{\tau_{LSS}-\tau}{\tau_{LSS}} \nabla_{\perp} \Phi  \, ,
\eeq 
where $\tau_O$ and $\tau_{LSS}$ denote respectively the conformal time at the observer and at the LSS; 
$\nabla_{\perp}$ stands for a gradient in the direction transverse to the line of sight.
In~\cite{MN2} we showed that the Lensing potential $\Theta$ can be written as $\Theta(\theta)=\Theta_0 p(\theta)$,
where $p(\theta)$ is a profile (see \cite{MN2} for its approximated expression) and 
$\Theta_0$ is the amplitude at the centre: 
\beq
\Theta_0 \approx \frac{1}{1.4} \, |\delta_0| \, (L H_0)^3 \frac{1}{D H_0} 
= \left(\frac{A ~L H_0 \tan^2\theta_L}{1-\frac{L H_0}{2\tan\theta_L}} \right)^{1/2} ~~~,
\label{tetazero}
\eeq
where the last equality follows by using eq.(\ref{eqdelta0}).

We show the dependence of $\Theta_0$ on $z$ in fig. \ref{figSNnd}.
Notice that when the Void is in the position closest to us, 
$\Theta_0$ reaches its minimum allowed value, which is about $3 \times 10^{-4}$. 
Clearly, we find this minimum
because we are imposing $A$ to be in the range suggested by present Cold Spot observations, 
$A=(7\pm 3)\times 10^{-5}$:
had we imposed a smaller value of $A$, we would have obtained a smaller minimum value for $\Theta_0$.

Given a temperature anisotropy $\Delta T^{(i)}(\hat {\bf n})/T$ (with $i=P,RS,L$) 
and the Lensing profile $\Theta(\hat {\bf n})$, 
their spherical harmonic decompositions, are respectively:
\beq
a_{\ell m}^{(i)} \equiv \int d \hat {\bf n}~ \frac{\Delta T^{(i)}(\hat {\bf n})}{T}~Y^*_{\ell m}(\hat{\bf n})~~~~ 
 ,~~~~~~~~ b_{\ell m} \equiv\int d \hat {\bf n}~ \Theta(\hat{\bf n}) ~ Y^*_{\ell m}(\hat {\bf n})~~~~.  
\label{almblm}
\eeq
Since the  profile is axially symmetric and since we have chosen 
the $\hat{z}$ axis to point towards the centre of the Void, 
the only non-vanishing $b_{\ell m}$ are those with $m=0$ and which, in addition, are real.

Given the $b_{\ell 0}$ coefficients in~(\ref{almblm}), we may compute the first order $a^{(L_1)}_{\ell m}$ 
coefficients for the Lensing temperature profile $\Delta T^{(L_1)}/T$, as in~\cite{MN2},
\beq
a_{\ell m}^{(L_1) }=\sum_{\ell', \ell''} G_{~\ell~~ \ell' \ell''}^{-m m 0}  
\frac{\ell'(\ell'+1)-\ell (\ell+1)+\ell''(\ell''+1)}{2} a^{(P)  *}_{\ell'-m}  b_{\ell'' 0}    ~~ ,
\label{almlensing}
\eeq
where we have introduced the Gaunt integrals, given in terms of the Wigner 3-j symbols~\cite{review} 
as follows:
\beq
 G_{\ell_1 ~\ell_2 ~\ell_3}^{m_1 m_2 m_3}\equiv \sqrt{\frac{(2 \ell_1+1)(2 \ell_2+1)(2 \ell_3+1)}{4 \pi}} 
 \left( \begin{matrix}  \ell_1 & \ell_2 & \ell_3 \cr 0 & 0 & 0 \end{matrix} \right)  
 \left( \begin{matrix}  \ell_1 & \ell_2 & \ell_3 \cr m_1 & m_2 & m_3 \end{matrix} \right) ~~ .
\eeq


\section{Non-diagonal Two-point functions}

For a primordial Gaussian signal the two-point correlation functions are given by:
\beq
\langle a^{(P)}_{\ell_1 m_1} a^{(P) \, *}_{\ell_2 m_2} \rangle=\delta_{\ell_1 \ell_2} 
\delta_{m_1 m_2} \langle C^{(P)}_{\ell_1} \rangle   \ ,
\label{infpred}
\eeq
where the $\langle C^{(P)}_\ell \rangle$ are predicted by some mechanism ({\it e.g.} inflation) that can 
generate primordial Gaussian fluctuations.

Given the expression in eq.(\ref{almlensing}), the first order contribution to the two-point 
correlation function due to the primordial and Lensing temperature fluctuations is:
\begin{eqnarray}
\langle a^{(P)}_{\ell_1 m_1} a^{{(L_1)}  *}_{\ell_2 m_2} \rangle 
&=& \sum_{\ell', \ell''} G_{~\ell_2 ~~\ell'~ \ell'' }^{-m_2 m_2 0 }  
\frac{\ell'(\ell'+1)-\ell_2 (\ell_2+1)+\ell''(\ell''+1)}{2} (-1)^{m_2}
\langle a^{(P)}_{\ell_1 m_1} a^{(P)  *}_{\ell' m_2} \rangle b_{\ell'' 0} \nonumber \\ 
&=& \delta_{m_1 m_2} (-1)^{m_2} \langle C^{(P)}_{\ell_1} \rangle \sum_{\ell''} G_{~\ell_2 ~~\ell_1 ~\ell''}^{-m_2 m_2 0}  
\frac{\ell_1(\ell_1+1)-\ell_2 (\ell_2+1)+\ell''(\ell''+1)}{2}   b_{\ell'' 0}  \, .
\label{2pointPL}
\end{eqnarray}
Note that this contribution is diagonal in the $m$ index. 
If one considers the diagonal in the $\ell$ index, $\ell_1=\ell_2$, this is a correction to the power spectrum. 
However, we have already shown in ~\cite{MN2} 
that - due to a property of the Gaunt integrals - such diagonal contribution vanishes.
In general there is also a contribution due to the coupling between the RS and Lensing temperature 
fluctuations, potentially inducing a non-vanishing effect on the diagonal. However, 
we have assumed that the presence of a Void in the direction of the Cold Spot is not correlated 
with the fluctuations of the LSS. 
Therefore such an effect is absent under our assumptions. 
In any case, even assuming a correlation, this would represent a subdominant contribution 
which does not change our results in an appreciable way.

Here we show that the non-diagonal terms are non-zero and measurable: in fact eq.(\ref{2pointPL}) leads 
to a correlation between different $\ell$'s. 
The correlations are small, but they are present also at high $\ell$'s. 
In fact any $\ell_1$ and $\ell_2$ will be correlated as long as $|\ell_1-\ell_2|\lesssim \Delta l$, 
where $\Delta l$ is a number between $25$ and $50$, depending on the chosen value for $\sigma$ 
(respectively $6^{\circ}< \sigma < 18^{\circ}$). This is because the Gaunt 
integral in eq.(\ref{2pointPL}) is non-zero if $|\ell_2-\ell_1|<\ell''$ while the coefficients $b_{\ell'' 0}$ 
are non-zero for $\ell''\lesssim \Delta l$.

Although the two-point function above is not invariant under rotations, it makes sense to consider its statistical 
average when decomposing the $a_{\ell m}$'s along the $\hat{z}$ axis directed towards the centre of the Cold Spot.
In order to be quantitative, we construct an estimator for the Signal-to-Noise ratio in the following way.
We define the quantities
\beq
F_{\ell_1 \ell_2 m} \equiv \frac{1}{2} \left(a^*_{\ell_1 m}~ a_{\ell_2 m} + a_{\ell_1 m}~ a^*_{\ell_2 m} \right)~~,
\eeq
whose average is given by
\beq
\langle F_{\ell_1 \ell_2 m} \rangle = 
\langle {a^{(P)}_{\ell_1 m}}^* a^{(L_1)}_{\ell_2 m} \rangle 
+ \langle {a^{(L_1)}_{\ell_1 m}}^* a^{(P)}_{\ell_2 m} \rangle~~,
\eeq
where in the last expression we neglected the quadratic contribution due to pure lensing. 
Due to the Gaunt integral property 
$G_{~\ell_1 ~\ell_2~ \ell'' }^{-m~ m ~0 } =G_{~\ell_2 ~\ell_1~ \ell'' }^{-m ~m~ 0 }$,
we have
\bea
\langle F_{\ell_1 \ell_2 m} \rangle=  (-1)^{m}  \sum_{\ell''}G_{~\ell~_1 \ell~_2 \ell'' }^{-m ~m ~0 } ~
\left( C_{\ell_1}^{(P)} ~\frac{\ell_1(\ell_1+1)-\ell_2 (\ell_2+1)+\ell''(\ell''+1)}{2} ~+ \right.~~~~~~~~ \nonumber \\ 
\left. + ~C_{\ell_2}^{(P)}~ \frac{\ell_2(\ell_2+1)-\ell_1 (\ell_1+1)+\ell''(\ell''+1)}{2} \right) ~ b_{\ell'' 0}~~~.
\eea
Since the variance of $\langle a^*_{\ell_1 m}~ a_{\ell_2 m} \rangle $ is 
$\sigma_F^2=\frac{1}{2}C_{\ell_1}^{(P)} C_{\ell_2}^{(P)} \left( 1+ \delta_{m 0}\right) $,
we can define a Signal-to-Noise ratio as:
\beq
\left( \frac{S}{N}\right)^2 = \sum_{\ell_1 \leq \ell_2,\ell_2 \leq {\ell_{\rm max}}, 0\leq m \leq \ell_1} \, 
\frac{\langle F_{\ell_1 \ell_2 m} \rangle^2}{\sigma_F^2} 
= \sum_{\ell_1 \leq \ell_2, \ell_2 \leq {\ell_{\rm max}},-\ell_1 \leq m\leq \ell_1 }  \, 
\frac{\langle F_{\ell_1 \ell_2 m} \rangle^2}{C_{\ell_1}^{(P)} C_{\ell_2}^{(P)} }  ~~ ,
\eeq
which is a function of $\ell_{\rm max}$.
Note that, since $F_{\ell_1 \ell_2 m}=F_{\ell_1 \ell_2 -m}$, in the first expression we summed only over 
$0\leq m \leq \ell_1$; however, in the last expression we have conveniently rewritten the sum over 
all $-\ell_1 \leq m \leq \ell_1$.

\begin{figure}[b!]\begin{center}
\includegraphics[width=7cm]{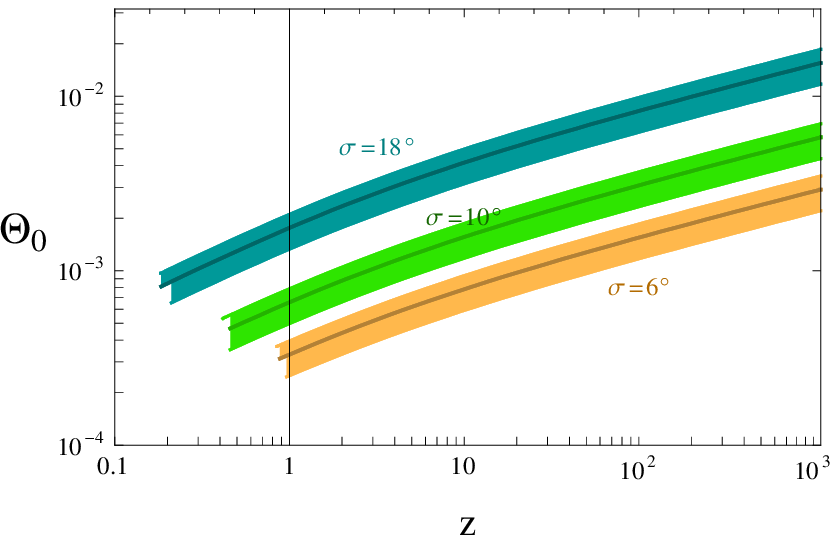} ~~~~~~~~~~~~~~
\includegraphics[width=7cm]{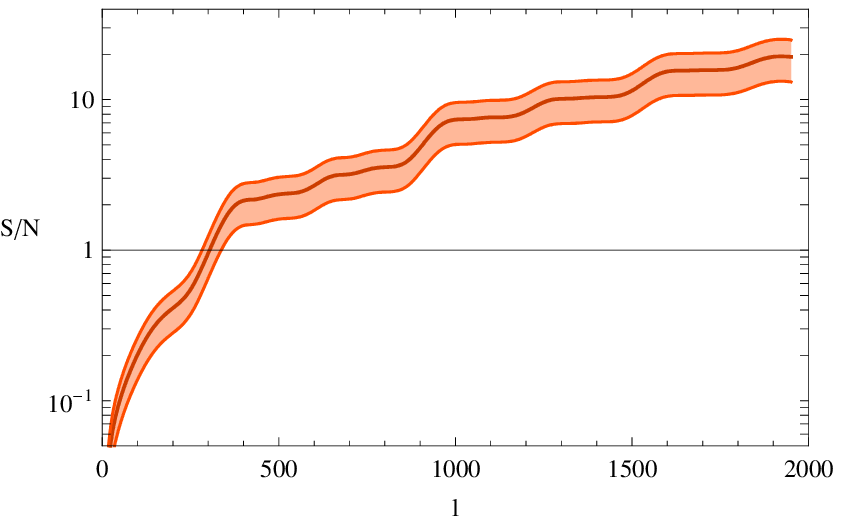}
\end{center}\vspace*{-0.5cm} 
\caption{Left: Plot of $\Theta_0$ as a function of the redshift $z$. 
The curves corresponds, as indicated, to $\sigma=6^\circ,10^\circ,18^\circ$. 
Right: Non-diagonal Signal-to-Noise ratio as a function of the multipole $\ell$, obtained 
by choosing the minimal value for $\Theta_0$ for any Void size (that is 
$\Theta_0 = (3,5,8) \times 10^{-4}$ for $\sigma = (6^\circ, 10^\circ,18^\circ)$ respectively) 
We recall that the ratio is linearly proportional to $\Theta_0$. 
In both plots, the shaded regions are obtained by varying $A$ in the range $(7\pm 3) \times 10^{-5}$.}
\label{figSNnd} 
\end{figure}

For each of the three Void sizes shown in the left panel of fig. \ref{figSNnd}, 
we calculate the Signal-to-Noise ratio as a function of $\ell_{\rm max}$ 
taking the corresponding minimal value of $\Theta_0$ (namely the weakest Lensing 
signal obtained when the Void is closest possible to the observer). 
These three curves actually overlap. The curve shown in the right panel of fig. \ref{figSNnd}
is then the minimum Signal-to-Noise ratio expected interpreting the Cold Spot as a Void.   
The plot can be easily adapted to larger values of $\Theta_0$, since the signal is just linearly proportional 
to $\Theta_0$, which can be read from the left panel of fig.\ref{figSNnd}. 
The result is that an experiment going up to $\ell_{\rm max} \sim 1000$, 
such as Planck (which should go further to about $2000$), should 
detect a signal for {\it any} Void size, with a Signal-to-Noise ratio larger than about 10.
In the WMAP data, which go up to $\ell\sim 700$, it could be already possible to find some signal, 
but it is not clear if the experimental noise  and the systematics will allow to see it for the entire 
parameter space, since the Signal-to-Noise ratio is lower.


\section{Conclusions}

Motivated by the so-called Cold Spot in the WMAP data, we have shown in this paper how to confirm 
the hypothesis that the Spot is due to an anomalously large Void along the line of sight.

In previous works we have analyzed rotationally invariant correlation functions, which lead to a detectable 
signal only in a fraction of the Void parameter space, because of special cancellations.
Here we have defined instead a 2-point function along the $\hat{z}$-axis, aligned towards the centre of the Spot, 
which is non-diagonal in $\ell$ space and does not suffer cancellations, leading to a much larger effect. 

We have shown that, for the whole Void parameter space, the Planck satellite should detect a clean signal
associated to such non-diagonal 2-point function, with a Signal-to-Noise ratio above 10.



\end{document}